\documentclass{PoS}
\usepackage{epsf}
\usepackage{graphicx}
\usepackage{amsmath}

\title{Semileptonic decays of  spin-1/2 doubly charmed baryons}

\ShortTitle{Semileptonic decays of  spin-1/2 doubly charmed baryons}

\author{\speaker{C. Albertus}\\
        Departamento de F\'\i sica Fundamental e
IUFFyM,\\ Universidad de Salamanca, E-37008 Salamanca, Spain\\
        E-mail: \email{albertus@usal.es}}

\author{E. Hern\'andez\\
        Departamento de F\'\i sica Fundamental e
IUFFyM,\\ Universidad de Salamanca, E-37008 Salamanca, Spain\\
        E-mail: \email{gajatee@usal.es}}

\author{J. Nieves\\
        Instituto de F\'\i sica Corpuscular
(IFIC), Centro Mixto CSIC-Universidad de Valencia, Institutos de
Investigaci\'on de Paterna, Aptd. 22085, E-46071 Valencia, Spain\\
        E-mail: \email{jmnieves@ific.uv.es}}

\abstract{We evaluate exclusive semileptonic decays of ground-state
  spin-1/2 doubly heavy charmed baryons. The decays are driven 
  by a $c\to s,d$  transition at the quark level. Our
  form factors are consistent with Heavy Quark Symmetry constraints. 
  The latter
   are valid in the limit of infinitely heavy quark mass at zero
  recoil.}

\FullConference{Sixth International Conference on Quarks and Nuclear Physics,\\
		April 16-20, 2012\\
		Ecole Polytechnique, Palaiseau, Paris}

\newcommand{\be}{\begin{equation}}
\newcommand{\ee}{\end{equation}}
\newcommand{\bea}{\begin{eqnarray}}
\newcommand{\eea}{\end{eqnarray}}

\makeatletter
\def\slashchar#1{{\mathpalette\c@ncel{#1}}} 
\makeatother

\begin{document}

\section{Decay width and form factor decomposition of the hadronic current}
The total decay width for semileptonic $c\to l$ transitions, with
$l=s,d$, is given by \bea \Gamma&=&|V_{cl}|^2
\frac{G_F^{\,2}}{8\pi^4}\frac{M'^2}{M} \int\sqrt{w^2-1}\, {\cal
L}^{\alpha\beta}(q) {\cal H}_{\alpha\beta}(P,P')\,dw \eea where
$|V_{cl}|$ is the modulus of the corresponding
Cabibbo--Kobayashi--Maskawa (CKM) matrix element for a $c\to l$ quark
transition, for which we shall use $|V_{cs}|=0.97345$ and
$|V_{cd}|=0.2252$ taken from Ref.~\cite{pdg10}.  $G_F= 1.16637(1)\times
10^{-11}$\,MeV$^{-2}$~\cite{pdg10} is the
Fermi decay constant, $P,M$ ($P',M'$) are the four-momentum and mass of
the initial (final) baryon, $q=P-P'$ and $w$ is the product of the
baryons four-velocities
$w=v\cdot v'=\frac{P}M \cdot \frac{P'}{M'}=\frac{M^2+M'^2-q^2}{2MM'}$. In the
decay, $w$ ranges from $w=1$, corresponding to zero recoil of the
final baryon, to a maximum value given, neglecting the neutrino mass,
by $w=w_{\rm max}= \frac{M^2 + M'^2-m^2}{2MM'}$, which depends on the
transition and where $m$ is the final charged lepton mass. Finally
${\cal L}^{\alpha\beta}(q)$ is the leptonic tensor after integrating
in the lepton momenta and ${\cal H}_{\alpha\beta}(P,P')$ is the
hadronic tensor.

The leptonic tensor is given by
\bea
{\cal L}^{\alpha\beta}(q)=A(q^2)\,g^{\alpha\beta}+
B(q^2)\,\frac{q^\alpha q^\beta}{q^2}
\eea
where
\bea
A(q^2)=-\frac{I(q^2)}{6}\left(2q^2-m^2-\frac{m^4}{q^2}\right)\ ,\ \ 
B(q^2)=\frac{I(q^2)}{3}\left({q^2+m^2}-2\frac{m^4}{q^2}\right)
\end{eqnarray}
 with $I(q^2)=\frac{\pi}{2q^2}(q^2-m^2)$.

The hadronic tensor reads
\begin{eqnarray}
{\cal H}^{\alpha\beta}(P,P') &=& \frac{1}{2J+1} \sum_{r,r'}  
 \big\langle B', r'\
\vec{P}^{\,\prime}\big| J_{cl}^\alpha(0)\big| B, r\ \vec{P}   \big\rangle 
\ \big\langle B', r'\ 
\vec{P}^{\,\prime}\big|J_{cl}^\beta(0) \big|  B, r\ \vec{P} \big\rangle^*
\label{eq:wmunu}
\end{eqnarray}
with $J$ the initial baryon spin, $\big|B, r\ \vec P\big\rangle\,
\left(\big|B', r'\ \vec{P}\,'\big\rangle\right)$ the initial (final)
baryon state with three-momentum $\vec P$ ($\vec{P}\,'$) and spin
third component $r$ ($r'$) in its center of mass frame. $J_{cl}^\mu(0)$ is
the charged weak current for a $c\to l$ quark transition \be
J_{cl}^\mu(0)=\bar\Psi_{l}(0)\gamma^\mu(1-\gamma_5)\Psi_c(0) \ee
Baryonic states are normalized as $ \big\langle B, r'\
\vec{P}'\, |\,B, r \ \vec{P} \big\rangle = 2E\,(2\pi)^3
\,\delta_{rr'}\, \delta^3 (\vec{P}-\vec{P}^{\,\prime}) $, with $E$
the baryon energy for three-momentum $\vec P$.

Hadronic matrix elements  can be parameterized in terms of form factors.
For $1/2 \to 1/2$ transitions the commonly used form factor decomposition reads
\begin{align}
\label{eq:1212}
&\hspace{3cm}\big\langle B'(1/2), r'\ \vec{P}^{\,\prime}\left|\,
\overline \Psi_l(0)\gamma^\mu(1-\gamma_5)\Psi_c(0)
 \right| B(1/2), r\ \vec{P}
\big\rangle =\nonumber\\ &{\bar u}^{B'}_{r'}(\vec{P}^{\,\prime})\Big\{
\gamma^\mu\left[F_1(w)-\gamma_5 G_1(w)\right]+ v^\mu\left[F_2(w)-\gamma_5
G_2(w)\right]
+v'^\mu\left[F_3(w)-\gamma_5 G_3(w)
\right]\Big\}u^{B}_r(\vec{P}\,) 
\end{align}

The $u_{r}$ are Dirac spinors normalized as $({ u}_{r'})^\dagger u_r =
2E\,\delta_{r r'}$. $v^\mu$, $v'^\mu $ are the four velocities of the
initial and final baryons. The three vector $F_1,\,F_2,\,F_3$ and
three axial $G_1,\,G_2,\,G_3$ form factors are functions of $w$ or
equivalently of $q^2$.\\

For $1/2 \to 3/2$ transitions we follow Llewellyn
 Smith~\cite{Llewellyn Smith:1971zm} to write
\begin{align}
\label{eq:1232}
&\hspace{0.8cm}\big\langle B'(3/2),r'\vec P'\,|\,\overline 
\Psi_l(0)\gamma^\mu(1-\gamma_5)\Psi_c(0)\,|\,B(1/2),r\,
\vec P\,\big\rangle=
~\bar{u}^{B'}_{\lambda\,r'}(\vec{P}\,')\,\Gamma^{\lambda\mu}(P,P')\,
u^{B}_r(\vec{P}\,)
\nonumber\\
&\hspace{2cm}\Gamma^{\lambda\mu}(P,P')=
\big[\frac{C_3^V(w)}{M}(g^{\lambda\,\mu}q
\hspace{-.15cm}/\,
-q^\lambda\gamma^\mu)+\frac{C_4^V(w)}{M^2}(g^{\lambda\,\mu}qP'-q^\lambda
P'^\mu)\\
&\hspace{4cm}+\frac{C_5^V(w)}{M^2}(g^{\lambda\,\mu}qP-q^\lambda
P^\mu)+C_6^V(w)g^{\lambda\,\mu}\big]\gamma_5\nonumber\\
&\hspace{0.5cm}+\left[\frac{C_3^A(w)}{M}(g^{\lambda\,\mu}q
\hspace{-.15cm}/\,
-q^\lambda\gamma^\mu)+\frac{C_4^A(w)}{M^2}(g^{\lambda\,\mu}qP'-q^\lambda
P'^\mu)+{C_5^A(w)}g^{\lambda\,\mu}+\frac{C_6^A(w)}{M^2}
q^\lambda q^\mu\right]\nonumber
\end{align}
Here $u^{B'}_{\lambda\,r'}$ is the Rarita-Schwinger spinor of the final spin
3/2 baryon normalized such that $(u_{\lambda\,r'}^{B'})^{\dagger}
u^{B'\,\lambda}_r = -2E'\,\delta_{rr'}$, and we have four vector
($C^V_{3,4,5,6}(w)$) and four axial ($C^A_{3,4,5,6}(w)$) form
factors.\\

\section{Heavy quark spin symmetry}
\label{sect:hqss}
In hadrons with a single heavy quark the dynamics of the light degrees
of freedom becomes independent of the heavy quark flavor and spin when
the mass of the heavy quark is much larger than $\Lambda_{QCD}$ and
the masses and momenta of the light quarks. This is the essence of
heavy quark symmetry (HQS)~\cite{hqs1,hqs2,hqs3,hqs4}. However, HQS
can not be directly applied to hadrons containing two heavy
quarks. The static theory for a system with two heavy quarks has
infra-red divergences which can be regulated by the kinetic energy
term $\bar h_Q (D^2/2 m_Q) h_Q$. This term breaks the heavy quark
flavor symmetry, but not the spin symmetry for each heavy quark
flavor~\cite{thacker91}. This is known as heavy quark spin symmetry
(HQSS). HQSS implies that all baryons with the same flavor
wave-function  are degenerate.  The invariance of
the effective Lagrangian under arbitrary spin rotations of the $c$
quark leads to relations, near the zero recoil point ($w=1
\leftrightarrow q^2=(M-M')^2 \leftrightarrow |\vec{q}\,|=0$), between
the form factors for vector and axial-vector currents between the
$\Xi_{cc}$ and $\Omega_{cc}$ baryons and the single charmed baryons. These decays are induced by the
semileptonic weak decay of the $c$ quark to a $d$ or a $s$ quark. The
consequences of spin symmetry for weak matrix elements can be derived
using the ``trace formalism''~\cite{Falk:1990yz}.

Near the zero recoil point, where the spin symmetry should
work best, HQSS considerably reduces the number of independent form
factors, and it relates those that correspond to transitions where the
spin of the two light quarks in the final baryon is $S=1$. Indeed, we
find at $w=1$ \cite{ahn11}
\begin{itemize}
\item $1/2\to 1/2$ transitions ($\Xi_{cc}\to \Lambda_c, \Xi_c$ and 
  $\Omega_{cc}\to \Xi_c$), where the total spin of the two light
quarks in the final baryon is $S=0$:
\begin{eqnarray}
 F_1+F_2+F_3=3G_1\equiv \eta_0
\label{eq:hqss1}
\end{eqnarray}
In the equal mass transition case one would find that $\eta_0$ is normalized as
$\eta_0(w=1)=\sqrt\frac32$. 
\item  Total spin of the two light quarks in the final
baryon is $S=1$ .
\begin{itemize}
\item[*]
$1/2\to 1/2$ transitions ($\Xi_{cc}\to \Sigma_c, \Xi'_c$ and 
  $\Omega_{cc}\to \Xi'_c, \Omega_c$) .
\begin{eqnarray}
F_1+F_2+F_3=\frac35G_1 \equiv \eta_1
\label{eq:hqss2}
\end{eqnarray}
\item[*] $1/2\to 3/2$ transitions ($\Xi_{cc}\to \Sigma^*_c, \Xi^*_c$ and 
  $\Omega_{cc}\to \Xi^*_c, \Omega^*_c$).
\begin{eqnarray}
\frac{\sqrt3}{2}\bigg(C_3^A\frac{M-M'}{M}+
C_4^A\frac{M'(M-M')}{M^2}+C_5^A\bigg) = \eta_1
\label{eq:hqss3}
\end{eqnarray}
\end{itemize}
In the equal mass transition case one would have that 
$\eta_1(w=1)=\frac1{\sqrt2}$ when
the two light quarks in the final state are different and
$\eta_1(w=1)=1$ when they are equal ($\Omega_c$ and $\Omega^*_c$).
\end{itemize}
Relations (\ref{eq:hqss1}), (\ref{eq:hqss2}) and (\ref{eq:hqss3}) are
exactly satisfied in the quark model when the heavy quark mass is made
arbitrarily large, and thus the calculation is consistent with HQSS
constraints.

\section{Results}
\begin{figure}[h!!!!]
\vspace*{.3cm}
\resizebox{4.8cm}{!}{\includegraphics{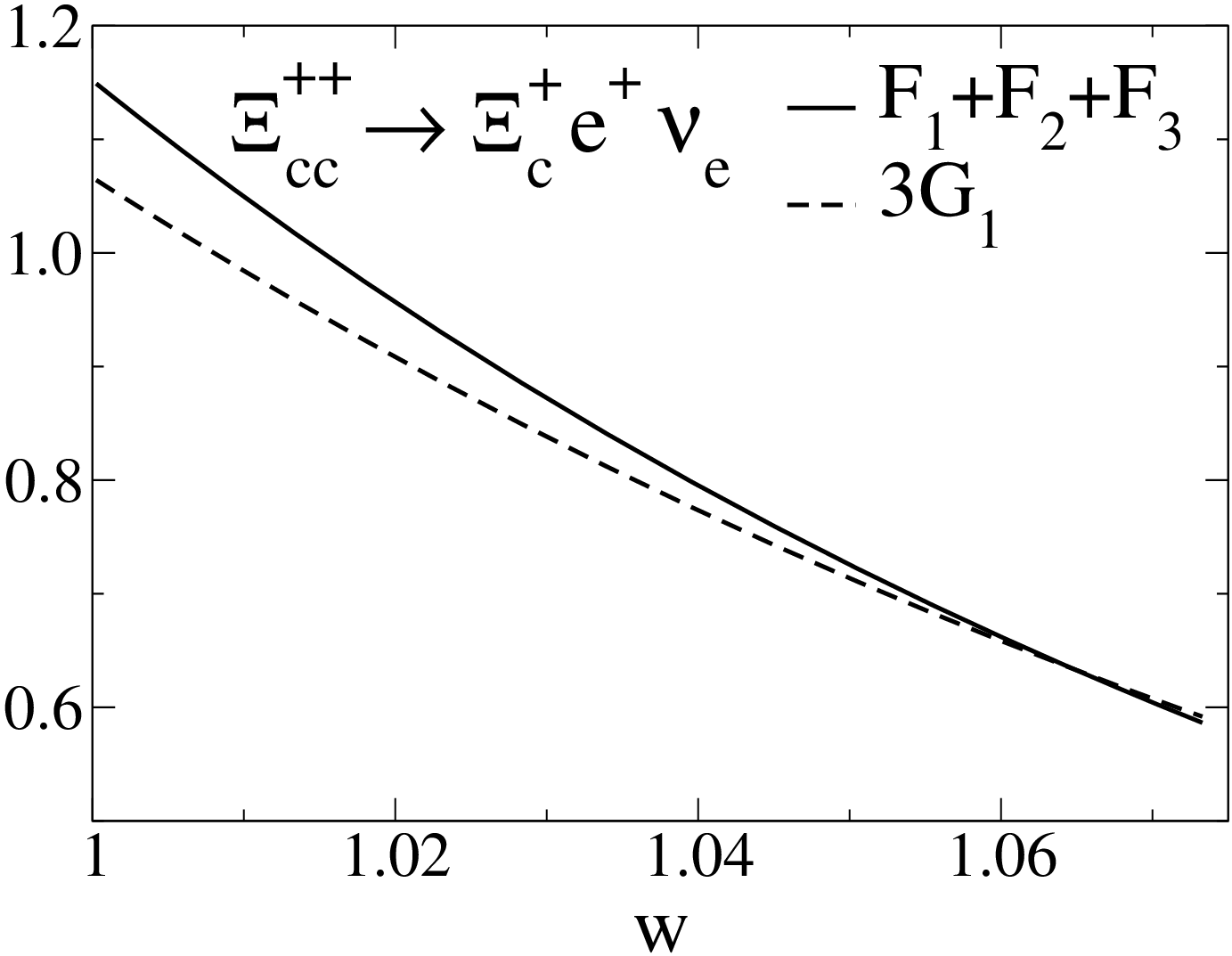}}\hspace{.25cm}
\resizebox{4.8cm}{!}{\includegraphics{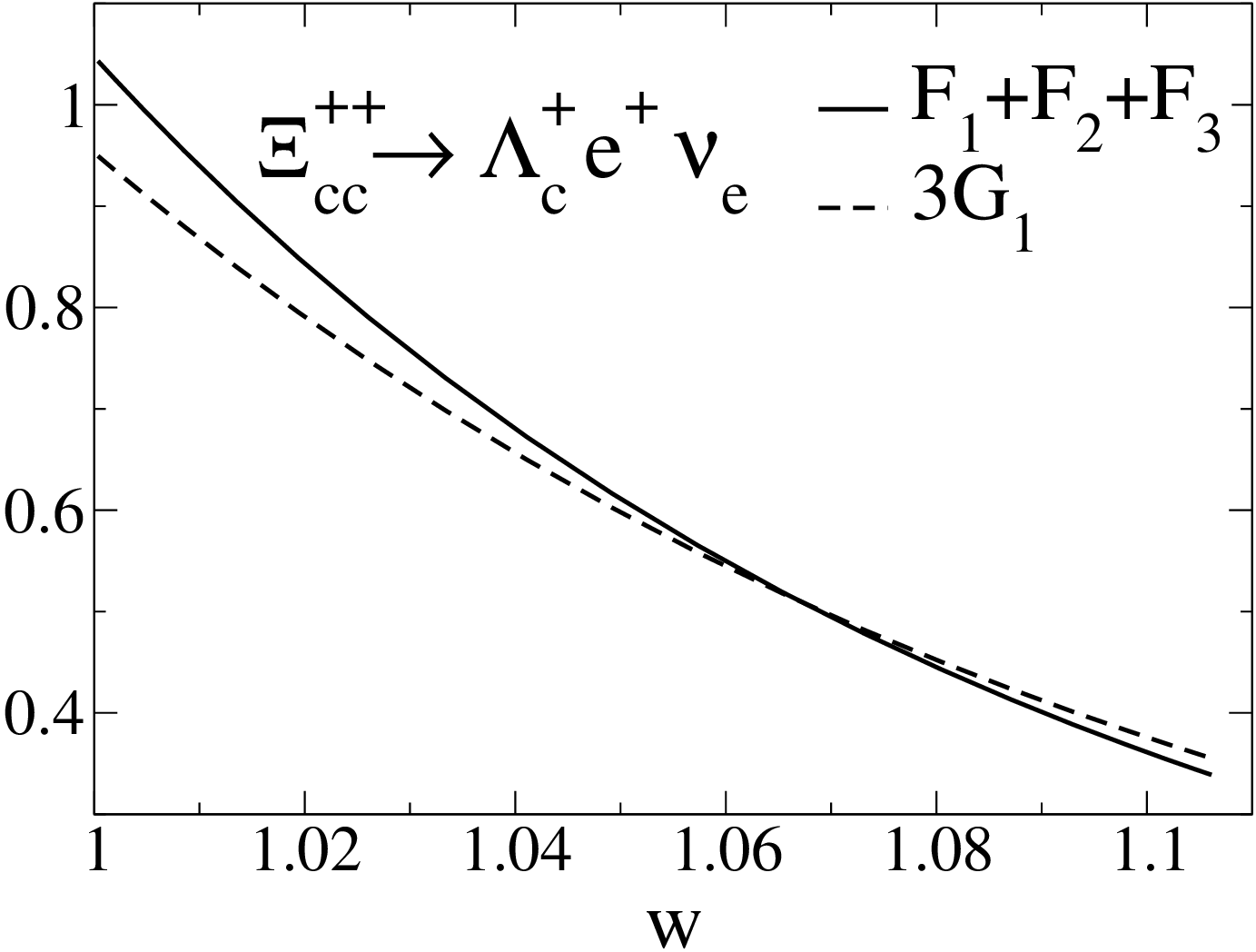}}
\hspace{.1cm}
\resizebox{4.8cm}{!}{\includegraphics{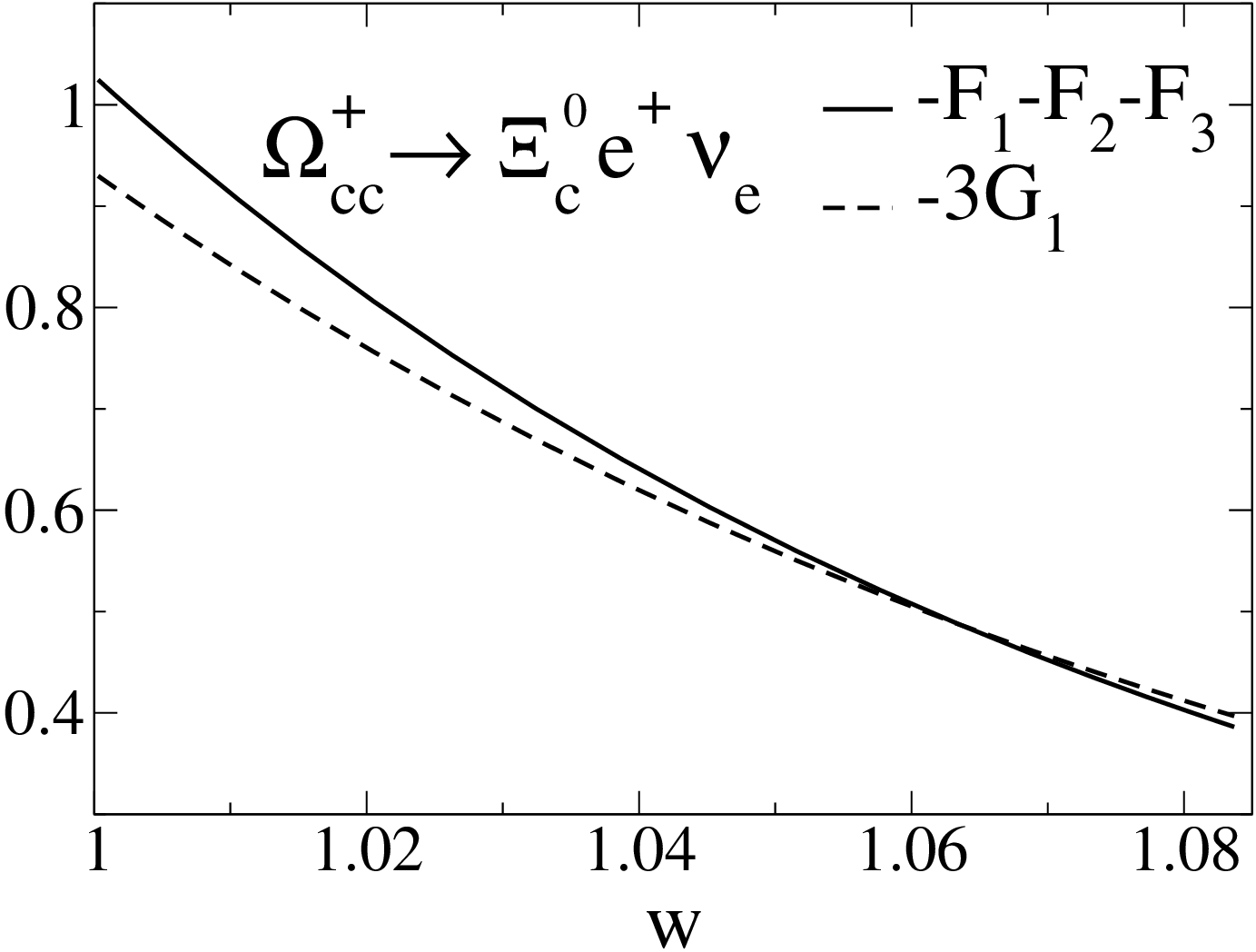}}
\caption{Comparison of $F_1+F_2+F_3$ (solid) and $3G_1$ (dashed) for the
specified transitions. The two light quarks in the final baryon have
total spin $S=0$. 
}\vspace*{.5cm}
\label{fig:s0}
\end{figure}

\begin{figure}[h!!!]
\begin{center}
\resizebox{5.35cm}{!}{\includegraphics{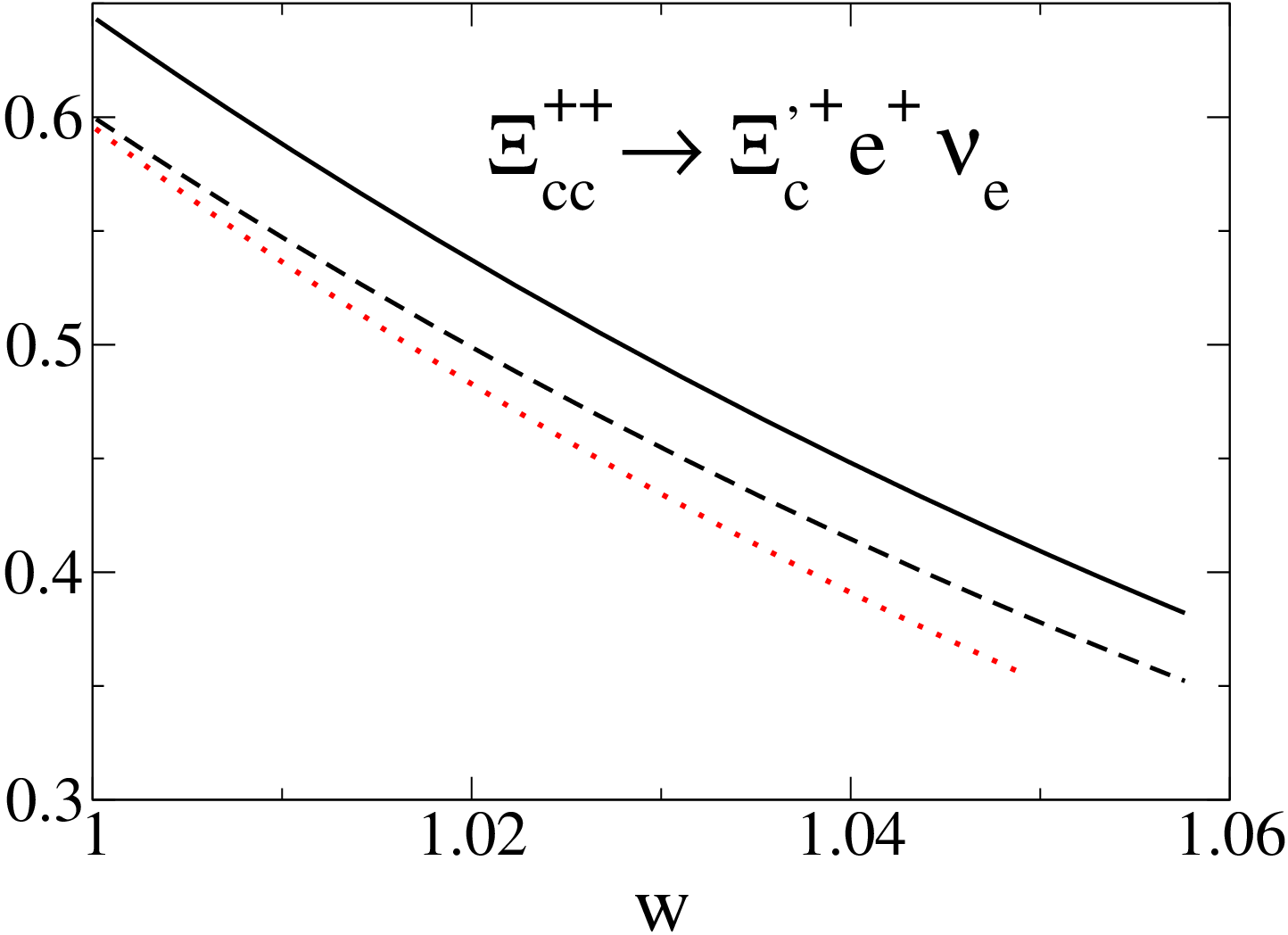}}\hspace{.75cm}
\resizebox{5.35cm}{!}{\includegraphics{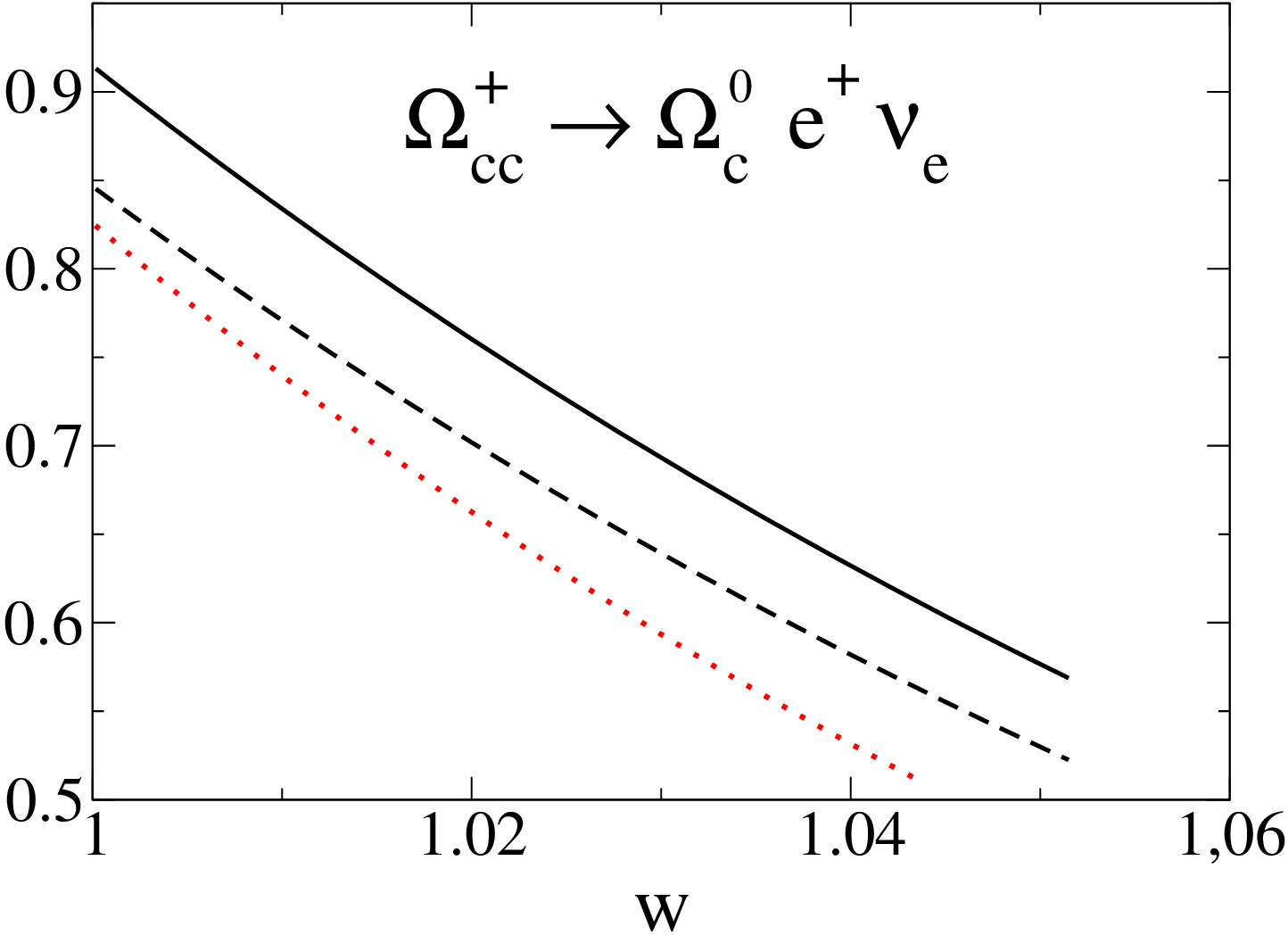}}\vspace{.7cm}\\
\hspace{-.1cm}
\resizebox{5.35cm}{!}{\includegraphics{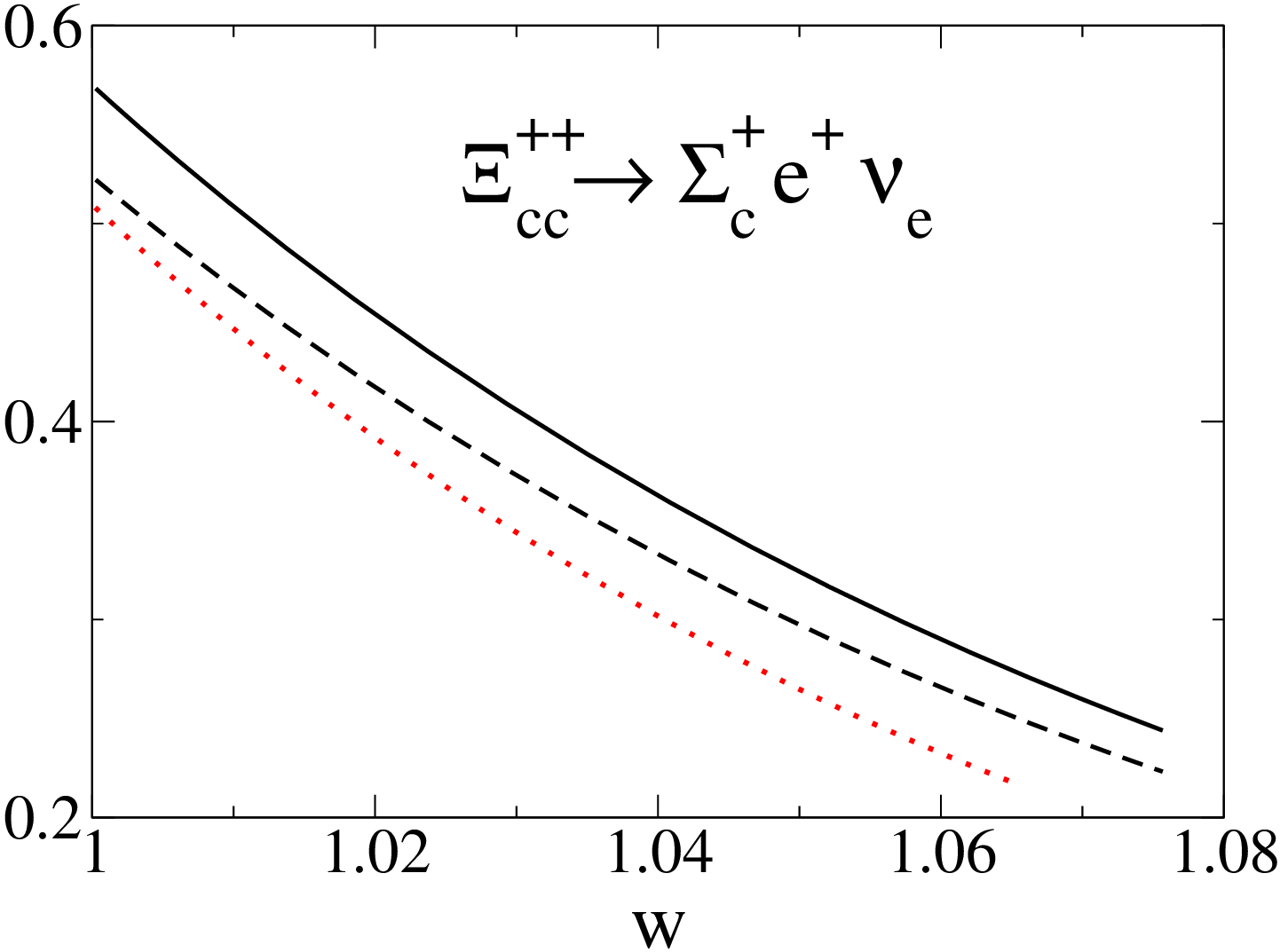}}
\hspace{.75cm}
\resizebox{5.1cm}{!}{\includegraphics{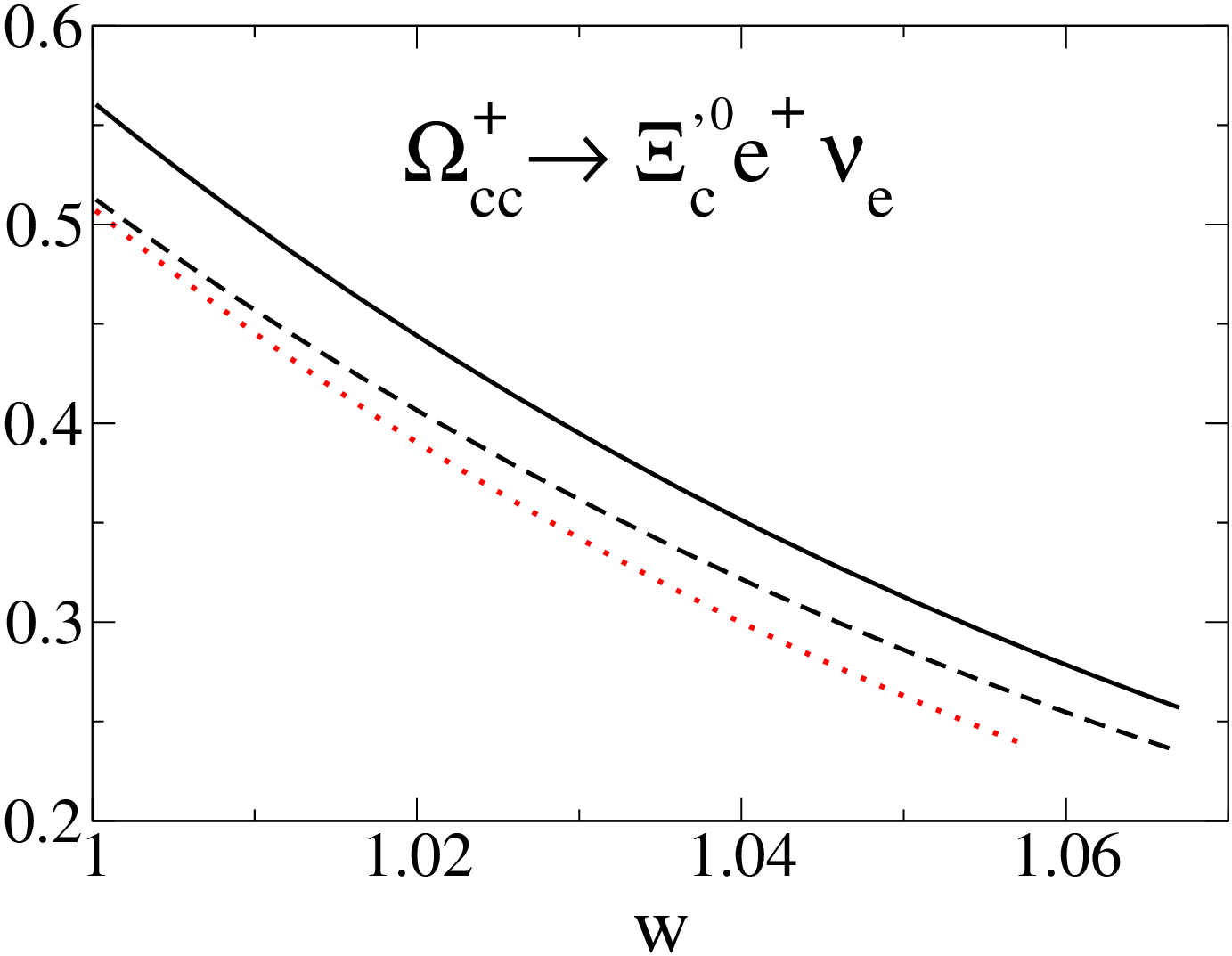}}
\caption{Solid (dashed): $F_1+F_2+F_3$ ($3G_1/5$) for the specified
transitions.  Dotted: the combination
$\frac{\sqrt3}{2}\big(C_3^A\frac{M-M'}{M}
+C_4^A\frac{M'(M-M')}{M^2}+C_5^A\big)$ for the transition with the
corresponding $3/2$ baryon ($\Sigma^*_c$, $\Xi^*_c$ or $\Omega^*_c$)
in the final state. In all cases the two light quarks in the final
baryon have total spin $S=1$. 
}
\label{fig:s1}
\end{center}
\end{figure}

We start by checking that our calculation respects the constraints on
the form factors deduced from HQSS.
 In Figs.~\ref{fig:s0} and \ref{fig:s1}, we show to what extent the
relations of (\ref{eq:hqss1}), (\ref{eq:hqss2}) and (\ref{eq:hqss3})
summarized above are satisfied for the actual $m_c$ value. In all cases
we see  moderate deviations, that stem from $1/m_c$ corrections, at the
level of  about 10\% near zero recoil, though larger than those
found in \cite{Hernandez:2007qv} for the $b\to c$ transitions of the
$\Xi_{bc}$ and $\Xi_{bb}$ baryons. These discrepancies tend to disappear
when the mass of the heavy quark is made arbitrarily large \cite{ahn11}.

%

%
Now we discuss the results for the decay widths. Those are shown in
Table~\ref{tab:dw} for the dominant ($c\to s$) and sub-dominant ($c\to d$)
exclusive semileptonic decays of the $\Xi_{cc}$ and $\Omega_{cc}$ to ground
state, $1/2^+$ or $3/2^+$, single charmed baryons and with a positron
in the final state\footnote{ Similar results are obtained for
$\mu^+\nu_\mu$ leptons in the final state.}.  For the $\Omega^+_{cc}$
baryon, semileptonic decays driven by a $s\to u$ transition at the
quark level are also possible.  However, in this latter case phase
space is very limited and we find the decay widths are orders of
magnitude smaller than the ones shown.  To our knowledge there are
just a few previous theoretical evaluations of the $\Xi_{cc}$ semileptonic
decays. In Ref.~\cite{Faessler:2001mr} the authors use the
relativistic three-quark model to evaluate the $\Xi_{cc}\to \Xi'_c
e^+\nu_e$ decay, while in Ref.~\cite{Kiselev:2001fw}, using heavy
quark effective theory and non-relativistic QCD sum rules, they give
both the lifetime of the $\Xi_{cc}$ baryon and the branching ratio for
the combined decay $\Xi_{cc}\to\Xi_c e^+\nu_e+\Xi'_c
e^+\nu_e+\Xi_c^*e^+\nu_e$ from which we have evaluated the
semileptonic decay widths shown in the table. We find a fair agreement
of our predictions with both calculations. In
Ref.~\cite{Guberina:1999mx}, using the optical theorem and the
operator product expansion, the authors evaluated the total
semileptonic decay rate finding it to be $0.151\,{\rm ps}^{-1}$ for
$\Xi_{cc}^{++}$ and $0.166\,{\rm ps}^{-1}$ for $\Xi_{cc}^{+}$. These
values are roughly a factor of two smaller than the sum of our partial
decay widths or the results in Ref.~\cite{Kiselev:2001fw}.  For the
$\Omega_{cc}^+$ a total semileptonic decay width of $0.454\,{\rm
ps}^{-1}$ is given in Ref.~\cite{Guberina:1999mx}. In this case this
is in better agreement with the sum of our partial semileptonic decay
widths which add up to $0.353\,{\rm ps}^{-1}$. 
\begin{table}
\begin{tabular}{lccc}\hline\hline
&&\multicolumn{2}{c}{ $\Gamma\ [\,{\rm ps}^{-1}]$} \vspace{.1cm}\\\cline{3-4}
$B_{cc}\to B_ce^+\nu_e$&\hspace*{.5cm}Quark transition\hspace*{.5cm} 
&This work&\cite{Faessler:2001mr},\cite{Kiselev:2001fw}\\\hline
$\Xi_{cc}^{++}\to\Xi^+_ce^+\nu_e$&$(c\to s)$&$8.75\times 10^{-2}$\\
$\Xi_{cc}^{+\hspace{.18cm}}\to\Xi^0_ce^+\nu_e$&$(c\to s)$&$8.68\times 10^{-2}$\\
$\Xi_{cc}^{++}\to\Xi'^+_ce^+\nu_e$&$(c\to s)$&0.146&$0.208\div 0.258$\cite{Faessler:2001mr}\\
$\Xi_{cc}^{+\hspace{.18cm}}\to\Xi'^0_ce^+\nu_e$&$(c\to s)$&0.145&$0.208\div 0.258$\cite{Faessler:2001mr}\\
$\Xi_{cc}^{++}\to\Xi^{*\,+}_ce^+\nu_e$&$(c\to s)$&$3.20\times 10^{-2}$\\
$\Xi_{cc}^{+\hspace{.18cm}}\to\Xi^{*\,0}_ce^+\nu_e$&$(c\to s)$&$3.20\times 10^{-2}$\\
$\Xi_{cc}^{++}\to\Xi'^+_ce^+\nu_e+\Xi^+_ce^+\nu_e+\Xi^{*\,+}_ce^+\nu_e$&
$(c\to s)$&0.266&$0.37\pm0.04^{(*)}$\cite{Kiselev:2001fw}\\
$\Xi_{cc}^{+\hspace{.18cm}}\to\Xi'^0_ce^+\nu_e\ +\Xi^0_ce^+\nu_e
\ +\Xi^{*\,0}_ce^+\nu_e$&$(c\to
s)$&0.264&$0.47\pm0.15^{(*)}$\cite{Kiselev:2001fw}\\
$\Xi_{cc}^{++}\to\Lambda^+_ce^+\nu_e$\hspace*{.5cm}&$(c\to d)$&$4.86\times
10^{-3}$\\
$\Xi_{cc}^{++}\to\Sigma^+_ce^+\nu_e$&$(c\to d)$&$7.94\times 10^{-3}$\\
$\Xi_{cc}^{+\hspace{.18cm} }\to\Sigma^0_ce^+\nu_e$&$(c\to d)$&$1.58\times 10^{-2}$\\
$\Xi_{cc}^{++}\to\Sigma^{*\,+}_ce^+\nu_e$&$(c\to d)$&$1.77\times 10^{-3}$\\
$\Xi_{cc}^{+\hspace{.18cm}}\to\Sigma^{*\,0}_ce^+\nu_e$&$(c\to d)$&$3.54\times
10^{-3}$\\
\hline
$\Omega_{cc}^{+\hspace{.18cm}}\to\Omega^{0}_ce^+\nu_e$&$(c\to s)$&0.282\\
$\Omega_{cc}^{+\hspace{.18cm}}\to\Omega^{*\,0}_ce^+\nu_e$&$(c\to s)$&$5.77\times 10^{-2}$\\
$\Omega_{cc}^{+\hspace{.18cm}}\to\Xi^{0}_ce^+\nu_e$&$(c\to d)$&$4.11\times
10^{-3}$\\
$\Omega_{cc}^{+\hspace{.18cm}}\to\Xi'^{0}_ce^+\nu_e$&$(c\to d)$&$7.44\times
10^{-3}$\\
$\Omega_{cc}^{+\hspace{.18cm}}\to\Xi^{*\,0}_ce^+\nu_e$&$(c\to d)$&$1.72\times
10^{-3}$\\
\hline\hline
\end{tabular}
\caption{Decay widths in units of ${\rm ps}^{-1}$. We use
$|V_{cs}|=0.97345$ and $|V_{cd}|=0.2252$ taken from Ref.~\cite{pdg10}. Results 
with an ${(\ast)}$, our
estimates from the total decay widths and branching ratios
in~\cite{Kiselev:2001fw}. Similar results are obtained for
$\mu^+\nu_\mu$ leptons in the final state.}
\label{tab:dw}
\end{table}
\begin{acknowledgments}
  This research was supported by DGI and FEDER funds, under contracts
   FPA2010-21750-C02-02, FIS2011-28853-C02-02, and the Spanish
  Consolider-Ingenio 2010 Programme CPAN (CSD2007-00042),  by Generalitat
  Valenciana under contract PROMETEO/20090090 and by the EU
  HadronPhysics2 project, grant agreement no. 227431. C. A. thanks a Juan de 
  la Cierva contract from the
Spanish  Ministerio de Educaci\'on y Ciencia.
\end{acknowledgments}

\end{document}